\documentstyle[aps]{revtex}

\input{epsf}
\def\dofig#1#2{\epsfysize=#1 \centerline{\epsfbox{#2}}}

\begin{document}
\title{Predictability of Currency Market Exchange}
\author{Toru Ohira\footnote{E-mail:ohira@csl.sony.co.jp}}
\address{
Sony Computer Science Laboratories\\
3-14-13 Higashi-gotanda, Shinagawa,\\
 Tokyo 141-0022, Japan\\
}
\author{Naoya Sazuka\footnote{E-mail:nsazuka@fe.dis.titech.ac.jp}}
\address{
Department of Computational Intelligence and Systems Science,\\
Tokyo Institute of Technology,\\
Yokohama 226-8502, Japan\\
}
\author{Kouhei Marumo\footnote{E-mail:kouhei.marumo@boj.or.jp}}
\address{
Institute for Monetary and Economic Studies\\
Bank of Japan\\
2-1-1 Hongoku-cho Nihonbashi, chuo-ku, Tokyo 103-8660, Japan\\
}
\author{Tokiko Shimizu\footnote{E-mail:tokiko.shimizu@boj.or.jp}}
\address{
Financial Markets Department\\
Bank of Japan\\
2-1-1 Hongoku-cho Nihonbashi, chuo-ku, Tokyo 103-8660, Japan\\
}
\author{Misako Takayasu\footnote{E-mail:takayasu@fun.ac.jp}}
\address{Future University-Hakodate\\
116-2 Kamedanakano-cho, 
Hakodate, Hokkaido,
Japan 041-8655\\
}
\author{Hideki Takayasu\footnote{E-mail:takayasu@csl.sony.co.jp}}
\address{
Sony Computer Science Laboratories\\
3-14-13 Higashi-gotanda, Shinagawa,\\
 Tokyo 141-0022, Japan\\
(Sony Computer Science Laboratory Technical Report: SCSL-TR-01-001)
}
\date{\today}
\maketitle
\begin{abstract}
We analyze tick data of yen-dollar exchange with a focus on its up and
down movement. We show that there exists a rather particular conditional
probability structure with such high frequency data. This result 
provides us with evidence to question one of the basic assumption 
of the traditional market theory, where such bias in high frequency
price movements is regarded as not present.
We also construct systematically a random walk model reflecting this probability
structure.
\end{abstract}
\vspace{2em}

One of the basic assumptions in the theory of economics is that the 
market is efficient and therefore it is not possible to exploit
or predict its behavior. A common
myth of traders, however, is that there is a certain predictability
in the market dynamics. There are several recent works in the field
of econophysics which indicate that these beliefs by traders are
real and the market in some aspect shows predictable dynamics\cite{zhang,lo,tsonis,johnson,takayasub}.
The main theme of this paper is to present another evidence of 
such predictability through an analysis of yen-dollar exchange
data. We use two sets of data composed of yen values
recorded at 7 second intervals on average.
We show that such "high frequency" data follows a rather 
particular conditional probability structure if we focus only 
on up or down movement. It may be impractical to
use this property for  actual trading given its high frequency and
existence of
transaction costs. However, it provides us with 
strong evidence to support an existence of probability structure
in high frequency price movements,
at least in certain markets.

Let us start our analysis with a description of the data sets.
We have obtained two data sets of yen-dollar exchange taken
for the period of 10/26/1998 to
11/30/1998 (data set A) and 1/4/1999 to 3/12/1999 (data set B) (Source:Bloomberg).
The time series data sets are composed of values $Y(t)$ of yen value at
"tick step" $t$. We note that $t$ is not real time, but rather discrete
steps with variable time intervals at which the exchange took place.
Data set A and B contain 267398 and 578509 data points, respectively. 
On average two ticks are separated by 7 seconds (Figures 1 (A) and (B)).
From these data sets we create a time series for the change  
$D(t) \equiv Y(t+1) - Y(t)$ for every $t$.
There are often cases where there is no change in $Y(t)$, ($D(t) =0$).
To make our analysis simpler, we will disregard such cases for the rest of this
paper. With this reduction, the data sets A and B contain 145542 and 344791 data points,
respectively, and the average time between ticks is about 10--13 seconds.
\begin{figure}[hbt]
\dofig{8cm}{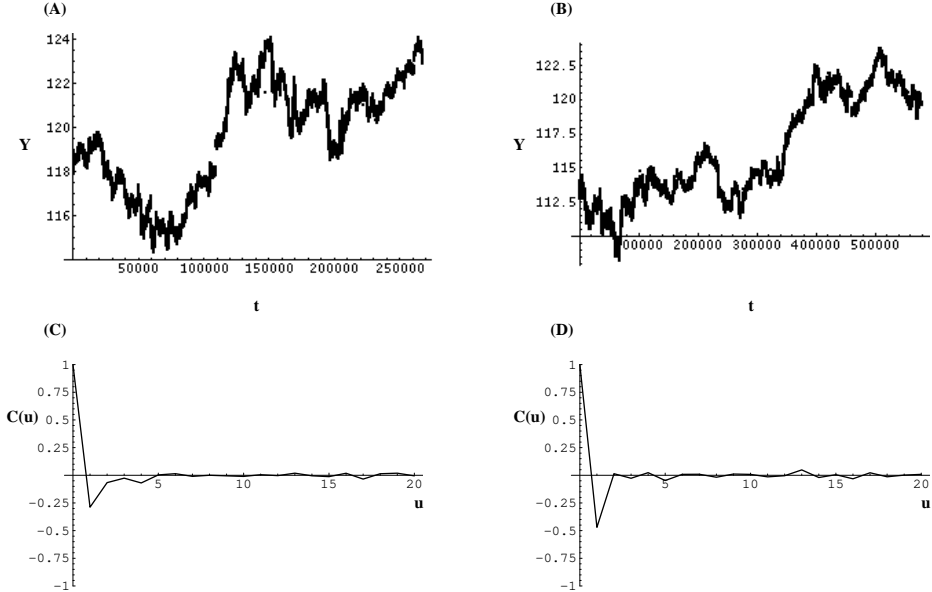}
\caption{
Time series plots of original data sets and the corresponding auto--correlation
function of their change at each tick. They are given as data set A in (A)(C) and 
data set B in (B)(D). 
} 
\label{fig1} 
\end{figure}

It has been observed\cite{takayasu} that the correlation function $C(u) \equiv <D(t+u)D(t)>$ computed from 
these data sets shows a negative correlation for a single tick (Figures 1(C) and (D)),
but almost zero for longer time intervals. This indicates that the direction of motion of $Y$ is 
more likely to be the opposite of that at the previous tick.
Though it is not realistic, let us assume that we can exploit this property at no cost.
Namely, we exchange our assets in accord with the following rule.  All our assets are changed into dollars
if the last tick of $Y(t)$ was down (yen appreciated), as we expect more chance for $Y(t)$ to increase
(dollar appreciated) in the next step. We exchange all our assets into yen in the opposite case.
To evaluate whether we can make a "profit" with this trading strategy, we calculate
the gain-loss function $g(k)$ for the asset A(t) using data set A:
\begin{equation}
g(k) = (A(t+k) - A(t))/A(t)
\end{equation}
This function represents the rate of gain or loss incurred with $k$ ticks apart.
The distribution function $Q(g(k))$ of the gain-loss function is plotted with different $k$
in Figure 2. We note that the peak of $Q(g(k))$ is moving in the positive direction indicating
more gain with larger $k$. Also, an asymmetry of distribution is seen.
To argue more quantitatively, we compute skewness \cite{smith}:
\begin{equation}
\rho = {{< (g - <g>)^3 >} \over {< (g -<g>)^2 >^{3/2}}}.
\end{equation}
The value of $\rho$ is $1.7 \sim 2.2$ for various $k < 2500$, reflecting on
the shape of the distribution $Q(g(k))$ being approximately similar with a positive skew.
This asymmetry with a positive skew is interpreted as being that this simple strategy enables one not only to
move along with the trend of the market, but also to gain more than
lose compared to the average trend. It should be stressed again that in reality
it is not possible to trade at every tick frequency and that there is a 
cost for each transaction. This observation, however, indicates that the high frequency yen-dollar
dynamics is not a mere random fluctuation around the overall trend, which may be affected
by other factors (such as interest rates) outside the market.
\begin{figure}[hbt]
\dofig{10cm}{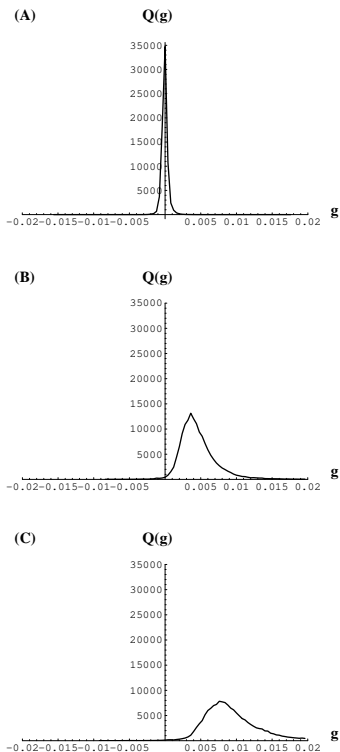}
\caption{
Distribution $Q(g(k))$ of the gain-loss function $g(k)$.
The values of $k$ are (A) $k=1$, (B) $k=64$, (C) $k = 128$.
} 
\label{fig2} 
\end{figure}

Motivated by this observation, we now extend our analysis of data 
beyond the correlation functions.
It will be shown that a highly common probabilistic structure is hidden
behind data sets A and B.
Based on this extended analysis we then propose a model which is a 
stochastic binary element whose
transition depends on its state at some preceding time steps.
We describe how the model can be constructed to
capture some statistical properties of the time series of
a real yen-dollar exchange.

From the data sets, we create a series $X(t)$ in the following way:
\begin{equation}
X(t) = +1 \quad (Y(t+1) - Y(t) > 0), \quad \quad X(t) = -1 \quad (Y(t+1) -
Y(t) < 0).
\end{equation}
In words, $X(t)$ reflects only information on $Y(t)$'s increase ($+1$)
or decrease ($-1$),
disregarding the amount of change (Figures 3 (A) and (B)). These are the
time
series we focus on for the rest of this letter. Now extracted Data A and B consists of
145542 and 344791 data points of $X(t)$, respectively.

The characteristics of $X(t)$ are more visible as we create a random walk,
$Z(t)$, by
$Z(t+1) = Z(t) + X(t)$ (Figures 3 (D) and (E) ).
 We note that the random walks are a mixture of one-directional and "zig-zag"
motion. These represent a mixture of trading "along" and "against" market trend 
as practiced by traders.  Also, we note
that
occasionally one-directional motion is rather persistent.  
This can be
seen
by the cumulative distribution $P(>t)$ of tick step length in the same direction
(Figures 3 (G) and (H)).  We see that around three steps the graph bends so
that the slopes get smaller for both data.  This indicates that after three
consecutive steps in the same direction there is more tendency for the
steps to continue to be in the same direction. This may be a reflection of dealer
psychology. If one directional moves last for a while, it is perceived as
a clear trend and more inclination is seen to "ride" with the market trend.

Let us now proceed with more quantitative analysis. 
We can compute from both data the conditional
probabilities of various orders for $X(t)$. These are summarized in
Table 1. We observe first that $X(t)$ is symmetric with respect
to positive ($+$) and negative ($-$) moves. Given this symmetry with
respect to $+$ and $-$, all the other conditional probabilities not
shown in the table can be derived from those shown.
We note the striking similarity of these values in Table 1 between
the two data sets, indicating the existence of a common probabilistic structure.
Zhang \cite{zhang} looked at similar higher order statistics
every half hour, and observed notable deviations for commodities like
silver but not for yen-dollar trading. At the tick level, there may exist a
very notable probabilistic rule as found in these data sets.

\begin{figure}[hbt]
\dofig{8cm}{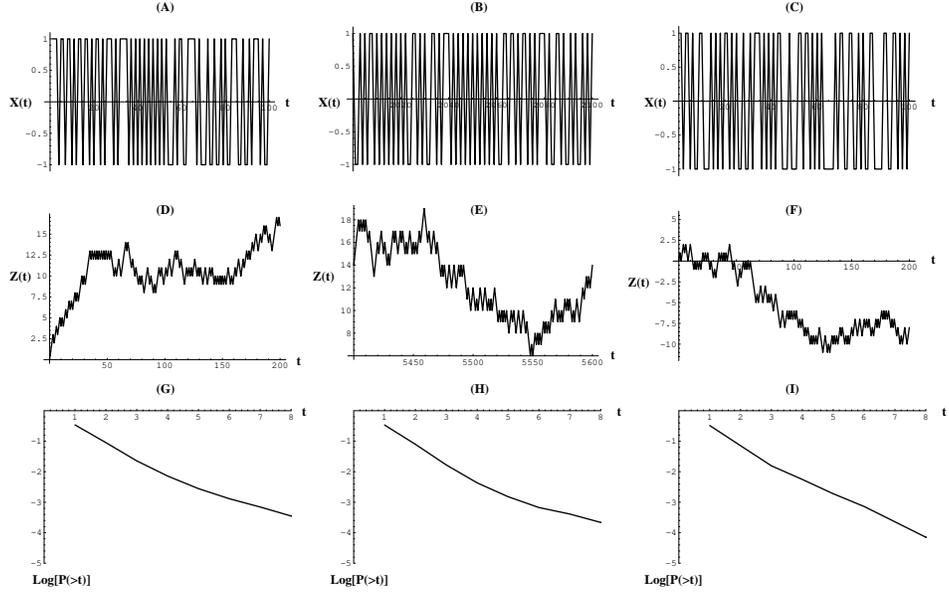}
\caption{
Sample time series plot $X(t)$ and $Z(t)$ and log plot of the cumulative
probability distribution function $P(>t)$. They are shown as data set A in (A)(D)(G), data set B 
in (B)(E)(H), and the model in (C)(F)(I).
} 
\label{fig3} 
\end{figure}

We now derive our model from close observation of Table 1.
With the assumption of the symmetry mentioned above, we
note that the second order conditional probabilities suffice to
explain most of the higher order values. The exception is
$P(+|+,+,+,+)$, which is notably higher than $P(+|+,+)$. Incorporation
of this correction leads to our model,  which is formally given
with three parameters:
\begin{equation}
P(+| +, +)  = p, \quad P(+|-,+) = q, \quad but \quad P(+|+,+,+,+) = p+w.
\end{equation}
We note that the derivation procedure presented above can iteratively
proceed to higher orders.  The model is presented here
with its simplicity valued. It can capture both qualitative and quantitative aspects of
data with a suitable choice of parameters as shown in 
Figures 3 (C), (F), (I),   and  Table 1.  (The parameters are 
$p = 0.22, q = 0.67, w = 0.15$. )   
Also, when we generate
a $\pm$ sequence from the model, the correct match with the data
is about $58\%$ for both data sets. 
On the other hand, the model cannot reproduce the concave up
feature of the $Log[P(>t)]$ distribution (Figures 3 (C), (F)) for $4 \leq t$, showing
a straight line (Figures 3 (I)).
One needs to incorporate further corrections to the model using higher order conditional
probabilities.

\begin{table}[hbt]
\dofig{7cm}{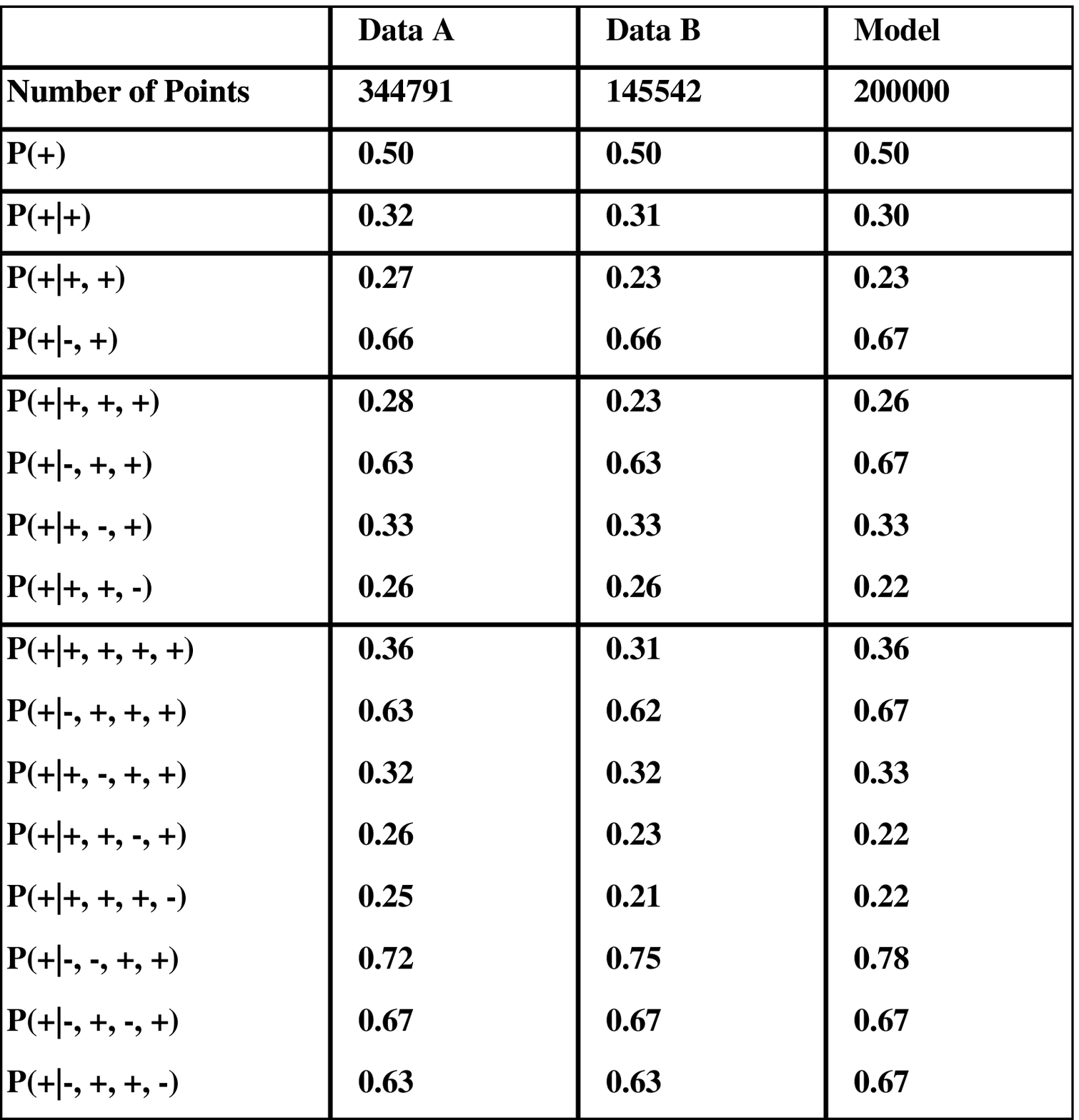}
\caption{
The values of joint probabilities computed from data sets A and B 
and from a simulation of the model.
The errors in these
numbers in the table is estimated to be around $0.01$.
} 
\label{table1} 
\end{table}

We may possibly design profitable strategies using higher order statistics of
the model. For example, we may only trade when expected probability deviations from 0.5 
are notable. Some preliminary results of such strategies
 are obtained which shows some "profit" even
with inclusion of transaction costs up to around 0.02 yen per dollar. 
This is still not enough to overcome actual cost, particularly when the
"ask-bid spread" is large.
More thorough investigation in this direction is left to the field of
financial engineering.

Our analysis of high frequency yen-dollar exchange data has indicated
 that such dynamics are not completely random and that a probabilistic
structure exists. Even though it is true that a trader can have great difficulty in
taking advantage of such a structure in reality, our investigation here
questions one of the basic assumptions of traditional economic theory. Together with the task of more thorough investigation with
wider samples of data sets, there are a series of questions to be studied.
They include such topics as correlations among price, volume and trading frequency,
and  market dynamics of other items, such as stocks, bonds, and so on.
Also, from the  point of view of random walks, our model given here is a variant of 
persistent and anti-persistent walks\cite{weiss}. It can also be viewed as an extension
of the stochastic binary element model with delay\cite{ohira-sato} or
random walk with delay\cite{ohira-yamane}. 
Theoretical studies of this model
are left for the future and may reveal interesting behaviors.
\vspace{2em}

One of the authors (T. O.) thanks Prof. J. D. Farmer of Santa Fe Institute for
his comments and suggestions on references.

\end{document}